\documentclass[twocolumn,journal]{IEEEtran}

\usepackage{amsmath}
\usepackage{amssymb}
\usepackage{graphicx}
\usepackage{hyperref}
\usepackage[numbers,square,sort]{natbib}
\usepackage{stmaryrd} 
\usepackage[labelfont=bf,font=footnotesize]{caption}
\usepackage{booktabs} 
\usepackage{setspace}

\newlength{\height}
\graphicspath{{figures/}}
\let\oldnameref\nameref
\renewcommand{\nameref}[1]{\textit{\oldnameref{#1}}}
\hypersetup{
    colorlinks=true,
    linkcolor=blue, 
    urlcolor=blue,
    citecolor=blue,
    pdftitle={Domain randomization},
}

\makeatletter
\renewcommand\@seccntformat[1]{}
\makeatother

\begin{document}

\title{
    Domain-randomized deep learning \\
    for neuroimage analysis\footnote{copyright}
}

\author{
    Malte Hoffmann
    \thanks{Malte Hoffmann (\url{mhoffmann@mgh.harvard.edu}) is with the Athinoula A.\ Martinos Center for Biomedical Imaging and the Departments of Radiology at Harvard Medical School and Massachusetts General Hospital.}
}

\markboth{IEEE Signal Processing Magazine,~accepted 15 July 2025}%
{Hoffmann: Domain-randomized deep learning for neuroimage analysis}

\IEEEpubid{
    \begin{minipage}{\textwidth}
    \vspace{0.5in}
    \centering
    \copyright~2025 IEEE. Personal use of this material is permitted. Permission from IEEE must be obtained for all other uses, in any current or future media, including reprinting/republishing this material for advertising or promotional purposes creating new collective works, for resale or redistribution to servers or lists, or reuse of any copyrighted component of this work in other works.
    \end{minipage}
}

\maketitle

\begin{abstract}
    Deep learning has revolutionized neuroimage analysis by delivering unprecedented speed and accuracy. However, the narrow scope of many training datasets constrains model robustness and generalizability. This challenge is particularly acute in magnetic resonance imaging (MRI), where image appearance varies widely across pulse sequences and scanner hardware. A recent domain-randomization strategy addresses the generalization problem by training deep neural networks on synthetic images with randomized intensities and anatomical content. By generating diverse data from anatomical segmentation maps, the approach enables models to accurately process image types unseen during training, without retraining or fine-tuning. It has demonstrated effectiveness across modalities including MRI, computed tomography, positron emission tomography, and optical coherence tomography, as well as beyond neuroimaging in ultrasound, electron and fluorescence microscopy, and X-ray microtomography. This tutorial paper reviews the principles, implementation, and potential of the synthesis-driven training paradigm. It highlights key benefits, such as improved generalization and resistance to overfitting, while discussing trade-offs such as increased computational demands. Finally, the article explores practical considerations for adopting the technique, aiming to accelerate the development of generalizable tools that make deep learning more accessible to domain experts without extensive computational resources or machine learning knowledge.
\end{abstract}


\section{Introduction}
\label{sec:intro}

\IEEEPARstart{N}{euroimaging} techniques, such as magnetic resonance imaging (MRI), have enabled the study of the human brain in vivo. Alongside advances in acquisition technology, research in neuroimage processing has led to software that automates systematic data analysis, minimizing human effort while improving accuracy and reproducibility~\cite{fischl2012freesurfer}.
In recent years, deep learning (DL) has been driving the development of a new class of algorithms with unprecedented speed and accuracy, and for a broad range of tasks, deep neural networks have largely replaced classical techniques.
However, a key challenge for DL in neuroimaging is small and highly specific datasets. Many studies include only hundreds or even tens of subjects~\cite{malone2013miriad}, due to factors such as the high cost of data acquisition, multiple modalities competing for scan time, the large size of multi-dimensional data like time-series acquisitions, the low prevalence of certain neurological disorders, and privacy concerns regarding data sharing~\cite{althnian2021impact}.
Training networks on limited datasets can lead to overfitting and poor generalization to new data---validation errors increase while the training loss continues to decrease~\cite{zhou2022domain}. This performance gap is common even for datasets acquired with similar MRI sequences in comparable cohorts, as models can become sensitive to subtle variations in scanner hardware or sequence parameters that trickle down to the images.

\begin{figure}
    \centering
    \includegraphics[width=\columnwidth]{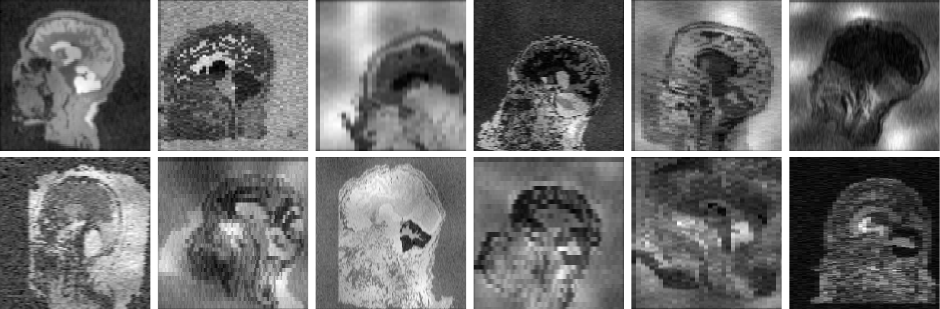}
    \caption{Synthetic training images. The variability intentionally exceeds realistic bounds of medical imaging to encourage deep neural networks to generalize. To realize the full potential of domain randomization, synthesis-driven training generates a new, unseen input image at every iteration.\label{fig:examples}}
\end{figure}

\begin{figure*}
    \centering
    \includegraphics[width=\textwidth]{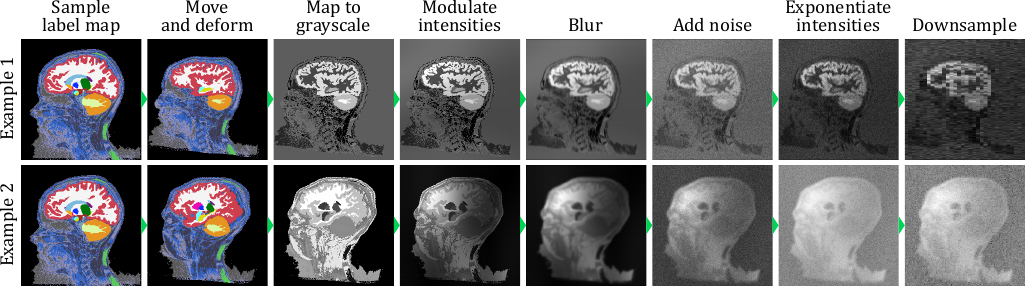}
    \caption{Image synthesis steps. First, we sample a previously generated anatomical label map, and randomly move and deform it. Second, we generate a grayscale image by drawing an intensity for each label. Third, a series of randomized image corruptions lead to complex intensity patterns across the image and each anatomical structure. Both rows begin with the same label map.\label{fig:synthesis}}
\end{figure*}

Emerging from a rich landscape of harmonization and domain shift mitigation techniques, a recent class of domain-randomization methods tackles the generalization problem by exposing networks to widely variable images synthesized from anatomical segmentation maps~\cite{gopinath2024synthetic} (Figure~\ref{fig:synthesis}). These methods address covariate shifts by generating an effectively unlimited stream of training data, varying key characteristics such as spatial structure, intensity, and resolution far beyond the realistic range (Figure~\ref{fig:examples}).
Synthesis-driven training is gaining traction in the neuroimaging community, as it facilitates the development of tools that generalize to new data types without retraining. It has been successfully applied to core neuroimage processing tasks such as segmentation~\cite{wells1996adaptive,van2003unifying} and registration~\cite{rueckert1999nonrigid}, which are fundamental to the interpretation of data acquired with a wide array of modalities.
However, training with synthetic data introduces an additional level of abstraction and new challenges, which can present a barrier to adoption. Departing from a recent survey of current and future applications of the strategy~\cite{gopinath2024synthetic}, this tutorial paper intends to serve as a practical guide to help new adopters navigate these challenges, select appropriate training strategies, and maximize the benefit of synthesis-driven training---to foster the development of robust, domain-invariant DL tools that empower users to analyze their data without DL expertise.

\IEEEpubidadjcol

\section{Background}
\label{sec:background}

A breadth of techniques aim to mitigate the effects of domain shift. This section provides a brief overview of domain shift and adaptation strategies. For details, we refer to more comprehensive reviews~\cite{csurka2017domain,wang2018deep}.

\subsection{Domain shift}
\label{sec:taxonomy}

Domain shift, or dataset shift, arises when the statistical distributions used for training and testing differ---common in neuroimaging and detrimental to machine-learning performance. Its three fundamental types are prior probability shift, concept shift, and covariate shift.

Denoting model inputs as $X$ and outputs as $Y$, prior shift occurs when the output distribution $P(Y)$ changes while the input characteristics given an output, $P(X | Y)$, remain the same. For example, a model trained on a balanced mix of infant and adult scans may underperform if tested predominantly on infant data---an effect often addressed by adjusting sampling probabilities during training.
Concept shift involves changes in the relationship between inputs and outputs, $P(Y | X)$. Imagine an age-prediction model that learns enlarged ventricles correlate with old age. If later deployed on pediatric Canavan disease patients, who also often have enlarged ventricles, the learned association breaks down.

Covariate shift occurs when the input distribution $P(X)$ changes while the input-output relationship $P(Y | X)$ remains unchanged. For instance, a brain segmentation model should ideally yield consistent results whether fed T1- or T2-weighted MRI. In practice, covariate shift arises from site effects (scanner differences) or batch effects (processing differences). Domain randomization targets both by diversifying $P(X)$ during training. The following sections review related strategies for mitigating covariate shift.

\subsection{Statistical harmonization}
\label{sec:harmonization}

Traditional harmonization corrects covariate shifts in neuroimaging studies by statistically adjusting the data. Empirical Bayes techniques like ComBat effectively remove batch effects estimated via linear models~\cite{johnson2007adjusting}, and extensions to this framework using generalized additive models allow for nonlinear effects.
These methods typically assume consistent preprocessing but require no labels and are computationally inexpensive.
Statistical moment matching reduces distributional discrepancies between a labeled source domain and an unlabeled or sparsely labeled target domain by reweighting source samples to better match target statistics. For example, kernel mean matching~\cite{gretton2009covariate} aligns features by minimizing maximum mean discrepancy (MMD) with a nonlinear kernel function, while correlation alignment matches second-order statistics for unsupervised domain adaptation~\cite{sun2017correlation}.

\subsection{Feature-based adaptation}

Feature-based, or shallow, domain adaptation estimates a transform that maps source data into the target domain via domain-invariant representations. A common approach constructs an intermediate low-dimensional space where source and target samples share features. For example, subspace alignment uses principal component analysis (PCA) to establish a linear mapping between source and target bases. Many other methods, such as geodesic flow kernels and transfer component analysis, refine this approach~\cite{csurka2017domain}.

\subsection{Deep domain adaptation}

More recently, DL has gained traction for domain adaptation. While early approaches use neural networks only for feature extraction followed by shallow adaptation, a vast landscape of modern deep methods integrate domain adaptation directly into representation learning using discrepancy metrics, adversarial learning, or reconstruction.

\subsubsection{Discrepancy-based adaptation}

Discrepancy-based domain adaptation fine-tunes model weights using target domain data. Early techniques align features from a single layer, while later approaches extend alignment across multiple layers. If target labels are available, discrepancy-based adaptation can use supervision. Otherwise, methods minimize statistical discrepancies using metrics like MMD (Section~\nameref{sec:harmonization}), directly regularize network weights to ensure a linear relationship between domains, or integrate adaptive batch normalization layers into the network.

\subsubsection{Adversarial domain adaptation}
\label{sec:adversarial}

Adversarial domain adaptation builds on generative adversarial networks (GANs), which adversarially train a generative and a discriminative model~\cite{pan2019recent}: the generator produces fake images to fool the discriminator, while the discriminator learns to distinguish real from synthetic examples. Generative adversarial adaptation follows this idea, generating simulated images in the target domain that remain compatible with source labels. In contrast, discriminative adaptation replaces the generator with a domain-invariant feature extractor, often involving image-to-image translation or style transfer to map data across domains. A common two-step approach first trains a feature extractor and task network on the source domain, then freezes their weights and adversarially trains a new extractor and discriminator on unlabeled target data~\cite{tzeng2017adversarial}.

\subsubsection{Reconstruction-based adaptation}

Reconstruction methods promote shared representations that support both the main task and image reconstruction, which is particularly useful when labels exist only in the source domain. Encoder-decoder networks, such as a variational auto-encoder (VAE), map inputs into a shared latent space and reconstruct them with a loss on the input. Adversarial variants introduce a domain confusion signal by training a discriminator to determine whether reconstructed samples come from the source or target domain. These methods tend to follow cyclic strategies, such as CycleGAN~\cite{wang2018deep}.

\subsection{Domain generalization}

Domain adaptation generally assumes access to source \textit{and} limited target domain data, where labels may exist for all target samples (supervised), some of the target samples (semi-supervised), or none of them (unsupervised)---the latter two being most common in neuroimaging. Domain adaptation bridges observed gaps given source and target samples, whereas domain generalization aims to improve model robustness across future, unseen domains, from which no samples are available at training time.

\subsubsection{Self-supervised learning}

Self-supervised learning replaces external labels with pretext tasks, leveraging unlabeled data for supervision. These tasks typically modify inputs to form related pairs and train models to predict relationships between them. Often used for pre-training, self-supervision can improve downstream performance when labeled data are scarce. A common approach is contrastive learning, which aligns representations of similar, ``positive'' pairs while separating representations of dissimilar, ``negative'' pairs.

\subsubsection{Data augmentation}

Data augmentation improves generalization by exposing models to more variability than the training set encompasses. Rule-based augmentation applies predefined transformations to images and label maps~\cite{shorten2019survey}. Common geometric transforms include flipping, affine, and elastic deformation, while intensity-based augmentation might add noise or stretch image histograms. More generic corruptions include swapping of image patches or convolutions with random kernels.
Learning-based augmentation trains networks to generate optimal transformations. Like \nameref{sec:adversarial}, adversarial augmentation uses a generator to create challenging transforms that fool a discriminator~\cite{pan2019recent}. In contrast, data-driven methods extract natural variations from auxiliary, unlabeled datasets, while uncertainty-guided augmentation learns transforms that target ambiguous inputs for which predictions are unreliable.

\subsubsection{Further approaches}

Several other domain generalization techniques extend performance to unseen settings without relying on target data during training. Multi-task learning trains a model to perform multiple related tasks at the same time to encourage general, shared representations useful for new tasks. Meta-learning is a related approach whose idea is to learn how to learn across a distribution of tasks. Ensemble learning trains multiple copies of the same model with varying initializations or training splits to improve robustness by fusing their predictions.

\subsection{Realistic simulations}

Synthetic images are widely used for DL and computer vision in neuroimaging~\cite{gopinath2024synthetic}, especially to address data scarcity with realistic simulations~\cite{shorten2019survey}. Instead of augmenting real data, these methods generate entirely new training images. Physics-based simulations use computational models to create medical images with controlled variability. VAEs synthesize anatomically plausible images by sampling from a learned latent space. Recently, probabilistic diffusion models have emerged as a powerful generative tool, producing diverse and detailed images that often surpass GANs in realism~\cite{muller2023multimodal}.

In contrast, domain randomization emphasizes data heterogeneity over authenticity to promote generalization across variations that may never be explicitly observed during training.

\begin{table*}
    \centering
    \caption{Meta-comparison of mean Dice-based accuracy rank percentiles across test sets for domain-randomization methods relative to state-of-the-art baselines. When not in top percentiles, domain-randomized methods typically exhibit only small performance gaps to the best-performing baseline---often a different method for each dataset. These datasets span structural MRI with T1-weighted (T1), T2-weighted (T2), proton-density-weighted (PD), and fluid-attenuated inversion recovery (FLAIR) contrast, MR angiography (MRA), diffusion-weighted images (DWI) and derived fractional anisotropy maps (FA), quantitative T1 maps (qT1), positron emission tomography (PET), computed tomography (CT), and optical coherence tomography (OCT).\label{tab:methods}}
    \small
    \begin{tabular}{llrrr}
    \toprule
              &                   & Baselines & Mean rank  & Mean Dice \\
    Main task & Modalities tested & tested    & percentile & gap to best \\
    \midrule
    Registration \\
    \midrule
    Affine$^{\,e}$~\cite{iglesias2023ready}
        & T1, T2, FA, FLAIR$^{\,s}$ & 2 & 71.9 & 0.3 \\
    Deformable$^{\,e}$~\cite{iglesias2023ready}
        & T1, T2, FA, FLAIR$^{\,s}$ & 3 & 54.2 & 1.3 \\
    Affine~\cite{hoffmann2024anatomy}
        & T1, T2, PD, post-contrast T1$^{\,c\,s}$ & 8 & 95.0 & 0.0 \\
    Deformable~\cite{hoffmann2024anatomy}
        & T1, T2, PD, post-contrast T1$^{\,c\,s}$ & 5 & 92.0 & 0.1 \\
    Longitudinal, rigid$^{\,e}$~\cite{fu2025learning}
        & T1, T2, post-contrast T1, FLAIR$^{\,c\,s}$ & 4--5 & 55.0 & 2.3 \\
    Longitudinal, rigid$^{\,w}$~\cite{fu2025learning}
        & T1, T2, post-contrast T1, FLAIR$^{\,c\,s}$ & 4 & 87.5 & 0.1 \\
    \midrule
    Segmentation \\
    \midrule
    Adult~\cite{billot2023synthseg}
        & T1, T2, PD, FLAIR, CT$^{\,s}$ & 3--6 & 92.3 & 0.3 \\
    Infant~\cite{hendrickson2023bibsnet}
        & T1, T2 & 1 & 100.0 & 0.0 \\
    Fetal$^{\,e}$~\cite{zalevskyi2024improving}
        & T2, T2$^{\,c}$ & 1--2 & 83.3 & 0.4 \\
    Adult~\cite{laso2024quantifying}
        & T1, FLAIR$^{\,c\,l\,s}$ & 1 & 100.0 & 0.0 \\
    Lesional~\cite{laso2024quantifying}
        & T1, FLAIR$^{\,c\,l\,s}$ & 1--2 & 100.0 & 0.0 \\
    Vascular~\cite{chollet2024neurovascular}
        & OCT & 1 & 100.0 & 0.0 \\
    \midrule
    Skull-stripping \\
    \midrule
    Adult~\cite{hoopes2022synthstrip}
        & T1, T2, PD, FLAIR, MRA, DWI, qT1, PET, CT$^{\,c\,s}$ & 5--6 & 98.0 & 0.0 \\
    Infant~\cite{kelley2024}
        & T1, T2 & 2--4 & 100.0 & 0.0 \\
    \bottomrule
    \multicolumn{5}{l}{
        $^{e\,}$Estimated from figures
        $^{w\,}$Whole head
        $^{s\,}$Include thick-slice stacks
        $^{c\,}$Include clinical data
        $^{l\,}$Include low-field MRI
    } \\
    \end{tabular}
\end{table*}

\section{Domain-randomized learning}
\label{sec:overview}

Domain randomization generates intentionally unrealistic training images from anatomical label maps, exposing networks to variability far beyond what limited real-world datasets typically capture.
Like other domain generalization strategies, it aims to promote generalization to domains unavailable during training. Domain randomization naturally integrates with supervised and semi-supervised learning. While it is compatible with self-supervised and unsupervised paradigms, we assume access to label maps, as these are the foundation for image synthesis.
Synthesis-driven training offers several advantages. First, it reduces the risk of overfitting~\cite{shorten2019survey}, as every training step presents the network with a new, unseen image.
Second, it can achieve state-of-the-art performance with only a few label maps, alleviating the need for compiling and annotating data.
Third, optimizing losses on select anatomical labels can produce anatomy-aware models. For example, registration networks can learn to align brains while ignoring other structures, such as the neck, which effectively eliminates the need for skull-stripping~\cite{hoffmann2021synthmorph}.
Fourth, labeling errors cannot degrade network performance, because the images are generated directly from label maps and thus match them exactly.
Finally, adding artifacts for a new modality or acquisition type is usually straightforward.

These factors have facilitated the development of methods that enable users to process their data without retraining or other techniques requiring DL expertise. Domain randomization has been successfully applied to structural segmentation~\cite{billot2023synthseg,hendrickson2023bibsnet,zalevskyi2024improving,laso2024quantifying,chollet2024neurovascular,dey2024anystar}, skull-stripping~\cite{hoopes2022synthstrip,kelley2024}, registration~\cite{hoffmann2021synthmorph,hoffmann2023anatomy,hoffmann2024anatomy,iglesias2023ready,fu2025learning}, feature extraction~\cite{dey2024learning}, image-to-image translation, and super-resolution reconstruction~\cite{iglesias2021joint}.

Table~\ref{tab:methods} presents a meta-comparison of a selection of these methods against state-of-the-art DL baselines trained with standard augmentation, as well as classical algorithms. For each study, we report the mean Dice-based accuracy rank percentile across the evaluated datasets, along with the mean Dice gap to the top-performing baseline---often a different baseline for each dataset. If there is a tie, we assign the average of the tied ranks.

Across a range of registration and segmentation tasks, the domain-randomization techniques generally achieve high rank percentiles. In cases of mid-range performance, the accuracy gap to the best-performing baseline is usually small. These findings suggest that while domain randomization does not always yield the highest accuracy, it generalizes well across diverse datasets---particularly heterogeneous clinical images, such as thick-slice acquisitions with glioblastoma---avoiding the gross inaccuracies that can occur with simpler methods~\cite{hoopes2022synthstrip,hoffmann2024anatomy}. We emphasize that some baselines do not support all datasets considered~\cite{billot2023synthseg,laso2024quantifying,fu2025learning}, which is not reflected in the percentile rankings of Table~\ref{tab:methods}. Domain-randomized runtimes are similar to conventional DL, typically substantially shorter than for classical algorithms.

Critically, domain randomization leverages a fully controllable, weightless generative model that synthesizes diverse anatomies and artifacts for training. The next section will review the key components of this model.
As MRI offers diverse contrast mechanisms and acquisition protocols, and is widely used in clinical and research neuroimaging due to its excellent soft-tissue contrast and non-invasive nature, many of these components derive from MRI acquisition. However, several studies have demonstrated the generalization of the presented modeling techniques to neuroimaging data acquired with computed tomography (CT)~\cite{hoopes2022synthstrip,billot2023synthseg,dey2024learning}, positron emission tomography (PET)~\cite{hoopes2022synthstrip}, and optical coherence tomography (OCT)~\cite{chollet2024neurovascular} (Table~\ref{tab:methods}). Beyond neuroimaging, these techniques have also been applied to 3D ultrasound, electron and fluorescence microscopy, and X-ray microtomography~\cite{dey2024learning,dey2024anystar}.

\section{Generative modeling}
\label{sec:modeling}

\def\N{\mathcal{N}}
\def\T{\mathcal{T}}
\def\U{\mathcal{U}}

While the implementation details of the generative model vary, the general concepts are the same across methods (Figure~\ref{fig:synthesis}).
We assume availability of a training set $\T$ of $N$-dimensional ($N$D, where $N$ is typically 2 or 3) anatomical label maps. The label maps could be synthetic~\cite{hoffmann2021synthmorph,chollet2024neurovascular}, derived from structural scans using various tools~\cite{billot2023synthseg,fischl2012freesurfer}, or sourced from public repositories~\cite{hoopes2022synthstrip}.
Let $s \in \T$ be a randomly selected label map and $g$ a generative model that receives $s$ as input and produces a new label map $s_x$ along with an associated, synthetic grayscale image $x$. In the absence of learnable parameters, $g$ uses simple physics-based and Gaussian-mixture modeling to generate $(x, s_x) = g(s, z)$ given random seed $z$, sampling synthesis parameters from uniform ($\U$) and normal distributions ($\N$) with zero mean.
In the following sections, we will break down the generative steps of Figure~\ref{fig:synthesis}. Some of these steps involve smoothly varying noise fields, which we will address first.

\setlength{\height}{0.16\textwidth}
\begin{figure*}
    \centering
    \begin{minipage}{0.28\textwidth}%
        \raggedright%
        \includegraphics[height=\height]{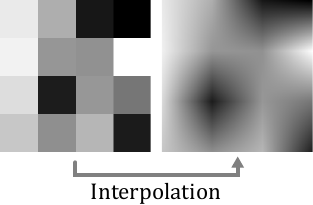}%
    \end{minipage}%
    \begin{minipage}{0.28\textwidth}%
        \centering%
        \includegraphics[height=\height]{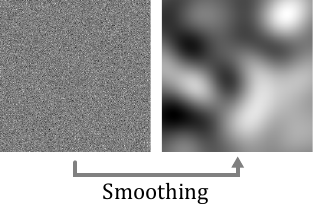}%
    \end{minipage}%
    \begin{minipage}{0.31\textwidth}%
        \raggedleft%
        \includegraphics[height=\height]{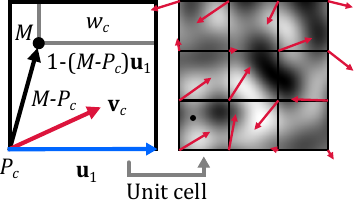}%
    \end{minipage}%
    \caption{Noise generation. \textbf{Left:} Linearly upsampling a random low-resolution image creates smoothly varying ``value noise'', which has a machine-generated appearance. \textbf{Center:} Gaussian noise, sampled at full resolution and smoothed via convolution, has a more natural appearance but is inefficient for large kernels. \textbf{Right:} Perlin noise---a type of gradient noise---achieves a natural look without convolutions. The intensity at each point is a combination of the dot products of random gradient vectors (red) at the corners of a unit cell and support vectors (black) from the same corners to that point.\label{fig:noise}}%
\end{figure*}

\subsection{Prerequisite: smooth noise}
\label{sec:noise}

Noise fields that smoothly vary across the spatial domain $\Omega$ are a prerequisite for spatial augmentation, intensity modulation, and label-map synthesis~\cite{hoffmann2021synthmorph}. As noise generation is ubiquitous in computer graphics, there are many efficient algorithms to choose from.

\subsubsection{Value noise}

A straightforward approach is to randomly sample a low-resolution field of size $f_1 \times f_2 \times \cdots \times f_N$, where we uniformly draw $f_i \sim \U(1, b_F)$, $f_i \in \mathbb{N}$ for each axis~$i \in \{1, 2, ..., N\}$, allowing the spatial frequency of the noise to vary across realizations. We then linearly upsample this field into $\Omega$.
The resulting image is called value noise and varies smoothly but has an artificial appearance (Figure~\ref{fig:noise}).

\subsubsection{Explicit smoothing}
\label{sec:explicit}

Alternatively, explicit smoothing results in more natural-looking noise and is particularly efficient for small Gaussian kernels when leveraging their separability to perform a series of 1D convolutions. First, we sample field $F \sim \N(\sigma_F ^ 2)$ of randomized standard deviation $\sigma_F \sim \U(a_F, b_F)$ at full resolution. Second, we construct a normalized Gaussian kernel $\kappa_{F,i}$ of uniformly sampled standard deviation $\sigma_{\kappa_F,i} \sim \U(a_{\kappa_F}, b_{\kappa_F})$ for each axis~$i$. We use kernels of length $|\kappa_{F,i}| = \lfloor 3 \times \sigma_{\kappa_F,i} \rceil \times 2 + 1$, where $\lfloor \cdot \rceil$ denotes rounding. Third, we convolve $F$ with these kernels, yielding
\begin{equation}
    F_\kappa = \alpha_F \times F \ast \kappa_{F,1} \ast \kappa_{F, 2} \ast \cdots \ast \kappa_{F, N},
\end{equation}
where rescaling by $\alpha_F = \max{\|F\|} / \max{\|F_\kappa\|}$ maintains the maximum strength of~$F$. We perform these convolutions using an $N$D routine, reshaping each 1D kernel into an $N$D kernel with singleton dimensions except along the axis~$i$.
Generating noise of low spatial frequency by smoothing via convolution is less efficient due to the need for larger kernels.

\subsubsection{Gradient noise}
\label{sec:gradient}

A third, widely used type of noise is gradient noise, which achieves a natural appearance without convolutions (Figure~\ref{fig:noise}). For example, Perlin noise procedures generate a smooth $N$D field $F$ by sampling gradient vectors of random orientation across a regular lattice of control points~\cite{perlin1985image}.
Let $P_c \in \mathbb{R}^N$ ($c \in \{1, 2, ..., 2^N\}$) be one of the $2^N$ control points defining a unit cell, $\mathbf{v}_c$ the gradient vector at $P_c$, and $\mathbf{u}_i$ the unit vector along spatial axis $i \in \{1, 2, ..., N\}$. We compute the field value at location $M$ within the cell as
\begin{equation}
    f(M) = \sum_{c=1}^{2^N}{w_c \times \mathbf{v}_c \cdot (M - P_c)},
\end{equation}
where $\cdot$ denotes the scalar product. The $N$D linear-interpolation weight (often faded)
\begin{equation}
    w_c = \prod_{i=1}^N{[1 - (M - P_c) \cdot \mathbf{u}_i]}
\end{equation}
ensures smooth transitions between control points. To vary the spatial frequency of the noise across realizations, we randomly sample the number of control points $C_i \sim \U(2, b_C)$, $C_i \in \mathbb{N}$ and process $C_i - 1$ unit cells along each axis~$i$.

Figure~\ref{fig:noise} compares $256 \times 256$ images of value noise generated by upsampling Gaussian noise from $1/64$ of the target resolution, normally distributed noise convolved with kernels of full width at half maximum (FWHM) 64, and Perlin noise on a gradient vector grid of $4 \times 4$ control points.

\subsubsection{Combining noise}

\begin{figure*}
    \centering
    \includegraphics[width=\textwidth]{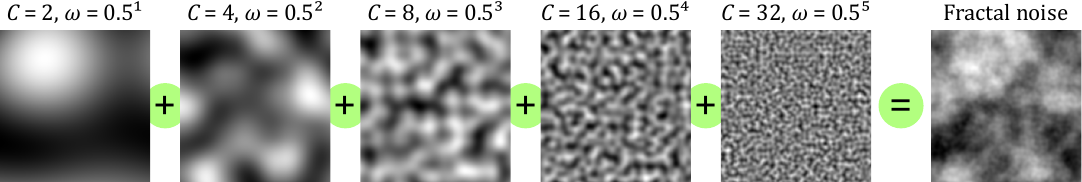}
    \caption{Fractal noise, or pink noise, results from adding noise over a range of spatial frequencies, with a relative weighting $\omega$ inversely proportional to the frequency. The example shown combines Perlin noise octaves sampled with $C \in \{2, 4, 8, 16, 32\}$ control points along each axis.\label{fig:fractal}}
\end{figure*}

We can create more complex noise patterns spanning a range of spatial frequencies by combining noise images generated at multiple resolutions. This type of noise is called fractal noise and typically involves halving or doubling the frequency at each level and summing these octaves with weights inversely proportional to the frequency. Figure~\ref{fig:fractal} illustrates this process with octaves of Perlin noise.
By design, Perlin noise has an intensity range of $[-1, 1]$. While we can directly control the intensity of value noise by adjusting the bounds of the uniform distribution, obtaining noise fields via smoothing requires sampling from a non-uniform distribution. For the following applications, we standardize the intensity range by min-max normalizing noise to the interval $[0, 1]$.
While the discussed techniques produce scalar fields, the first modeling step of Figure~\ref{fig:synthesis}---spatial augmentation---requires a vector field for deformation, which we generate by sampling $N$ independent noise components.

\subsection{Spatial augmentation}
\label{sec:augmentation}

This step aims to simulate variations in head orientation within the field of view (FOV) by translating and rotating the label map $s$. Additionally, smooth, nonlinear transformations including scaling and shear increase anatomical variability. Let $\mathbf{\Phi} = \mathbf{\phi} \circ A$ denote the composition of an affine transformation $A$ and a nonlinear deformation field $\mathbf{\phi}$.

\subsubsection{Affine transformation}
\label{sec:affine}

We generate $A$ in 3D as follows. For each axis~$i$, we sample parameters for translation $t_i \sim \U(a_t, b_t)$, rotation $r_i \sim \U(a_r, b_r)$, scaling $z_i \sim \U(a_z, b_z)$, and shear $e_i \sim \U(a_e, b_e)$. From these, we construct the corresponding $4 \times 4$ matrix transforms $T$, $R$, $Z$, and $E$, respectively, which we compose as $A = T \circ R \circ Z \circ E$. The simpler 2D case uses two parameters for translation and scaling each, and one parameter for rotation and shear each, resulting in a $3 \times 3$ matrix transform~\cite{hoffmann2024anatomy}.

\subsubsection{Elastic deformation}

The nonlinear component $\mathbf{\phi}$ is a vector field. We generate it by sampling a min-max normalized $N$-component field $\hat{\mathbf{\phi}}$ using one of the methods from Section~\nameref{sec:noise} and by randomly scaling this field to obtain $\mathbf{\phi} = \|\mathbf{\phi}\| \times \hat{\mathbf{\phi}}$, where $\|\mathbf{\phi}\| \sim \U(a_\mathbf{\phi}, b_\mathbf{\phi})$.
To ensure that $\mathbf{\phi}$ does not introduce holes or folding, we can similarly generate a smooth, stationary velocity field (SVF) $\mathbf{\nu}$ instead and integrate it over unit time to obtain a diffeomorphism~\cite{arsigny2006log}.

Applying the composite transform $\mathbf{\Phi} = \mathbf{\phi} \circ A$ to the label map $s$ using nearest neighbor interpolation yields a new label map, $s_x = s \circ \mathbf{\Phi}$.

\subsubsection{Partial field of view}
\label{sec:mask}

Often, brain scans do not capture the full anatomy to save time. We simulate partial-FOV acquisitions by cropping the image content. From $s_x$, we derive a binary cropping mask $m$ that zeroes out a proportion $p_m \sim \U(a_m, b_m)$ of the outermost voxels along a random axis. An efficient implementation is to construct 1D binary masks $m_i$ along each axis $i$, reshape them to have singleton dimensions along all other spatial axes, and combine them into the final mask $m$ via element-wise multiplication using the broadcasting mechanics of modern DL libraries:
\begin{equation}
    m = m_1 \odot m_2 \odot \cdots \odot m_N.
\end{equation}

Internally, model $g$ generates image $x$ from the label map $s_x$ (Figure~\ref{fig:synthesis}). Depending on the target task, $g$ may return $s_x$ or the cropped label map $s_x \odot m$ (Figure~\ref{fig:synthesis}). For example, a segmentation model may label structures in $s_x \odot m$ that are present in a cropped image, whereas an affine registration model could learn to align all structures in $s_x$ even if $x$ has a partial FOV.
The next step is synthesizing a grayscale image.

\subsection{Image synthesis}
\label{sec:synthesis}

The image synthesis builds on a Bayesian model of MRI contrast~\cite{wells1996adaptive}, which assumes that the voxel intensities within each anatomical structure $j$ in $s_x$ follow a Gaussian distribution. However, the noise level is unlikely to vary across tissue types, and we do not want to provide this statistic to the downstream task network. In the absence of artifacts, we therefore treat image voxels associated with label $j$ as independent samples from a normal distribution $\N(\mu_j, \sigma_n^2)$ with label-specific mean $\mu_j$ and global variance $\sigma_n^2$.

We generate a noise-free ``mean'' intensity image $x_\mu$ by assigning a random intensity value $\mu_j \sim \U(0, 1)$ to all voxels with label $j$. As a result, the left hippocampus might be bright and the right hippocampus dark in one batch, and this contrast may reverse in the next. Although we could constrain bilateral intensities to be the same or use per-label intensity ranges to simulate a specific modality such as CT, these constraints can limit generalization. 
Counterintuitively, more realistic synthesis rules have been shown to underperform relative to unconstrained sampling even for the target modality~\cite{billot2023synthseg}.

Indexing into a lookup table is an efficient approach to implementing a function $\mu : \llbracket 0, J - 1 \rrbracket \longrightarrow [0, 1]$ that associates zero-based index labels $j \in \{0, 1, ..., J - 1\}$ with the $J$ intensity values $\mu_j$. We sample $\mu$ as a 1D tensor and compute the mean image $x_\mu$ by treating $s_x$ as an index map into $\mu$, assigning each voxel location $M \in \Omega$ the intensity $x_\mu(M) = \mu(s_x(M))$.
Next, we will corrupt the image.

\subsection{Image corruptions}

Applying a series of randomized corruptions to the noise-free image $x_\mu$ will create complex intensity patterns across each anatomical structure (Figure~\ref{fig:synthesis}). As variations in noise levels across space are undesirable, we modulate image intensities with a bias field \textit{before} adding noise.

\subsubsection{Bias field}

A common MRI artifact is a low-frequency intensity non-uniformity of 10--20\% across the image, often referred to as an intensity bias~\cite{sled1998nonparametric}. This effect arises from multiple factors, including eddy currents induced by gradient field changes and non-uniformities in the radio-frequency coils. In order to simulate an intensity bias, we generate a smooth scalar noise field $\hat{B}$, normalized into the $[0, 1]$ range (Section~\nameref{sec:noise}), and apply it multiplicatively to the image:
\begin{equation}
    x_B = x_\mu \odot (id - \|B\| \times \hat{B}),
\end{equation}
where $id$ is the identity field, and $\|B\| \sim \U(a_B, b_B)$ is the maximum intensity drop, sampled uniformly.

\begin{figure}
    \centering
    \includegraphics[width=0.7\columnwidth]{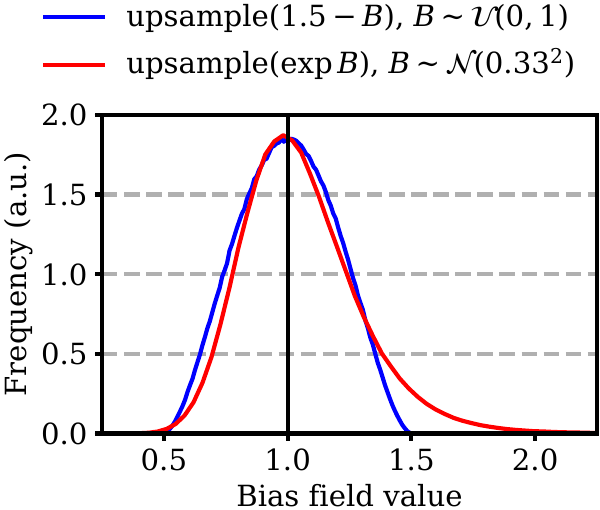}
    \caption{Bias field distribution after upsampling. Uniform sampling results in a bounded, symmetric distribution (blue, shifted by 1.5 for comparison). In contrast, applying the exponential function to normally distributed values guarantees a positive field, but the resulting distribution is asymmetric and includes higher values (exceeding 2 for standard deviation $\sigma = 0.33$, red).\label{fig:bias}}
\end{figure}

An alternative implementation~\cite{billot2023synthseg,hoopes2022synthstrip} samples a low-resolution bias field from a normal distribution without re-normalizing intensities, applies the exponential function voxel-wise to map values into $\mathbb{R}^+$, and then linearly upsamples the field into $\Omega$. As shown in Figure~\ref{fig:bias}, this approach does not modulate intensities symmetrically, which may be undesirable.

\subsubsection{Gaussian blur}
\label{sec:blur}

Similar to the bias field, we apply random blur \textit{before} adding noise to ensure a diverse landscape of images---adding noise first would correlate the noise floor with the smoothness level. Blurring simulates partial volume effects, where voxel intensities are a combination of signals from different tissue types due to the finite voxel size~\cite{van2003unifying}.

As in Section~\nameref{sec:noise}, we construct a normalized Gaussian kernel $\kappa_i$ of uniformly sampled standard deviation $\sigma_{\kappa,i} \sim \U(a_{\kappa}, b_{\kappa})$ for each axis $i \in \{1, 2, ..., N\}$. We convolve $x_B$ with these kernels, yielding
\begin{equation}
    x_\kappa = x_B \ast \kappa_1 \ast \kappa_2 \ast \cdots \ast \kappa_N.
\end{equation}

\subsubsection{Additive noise}

The unstructured background noise in MRI magnitude images follows a Rician distribution and primarily arises from electrical resistance. However, for simplicity, we corrupt the image with Gaussian noise (Section~\nameref{sec:synthesis}). We uniformly sample a standard deviation $\sigma_n \sim \U(a_n, b_n)$ and draw a Gaussian noise tensor $n \sim \N(\sigma_n^2)$ across $\Omega$ to obtain $x_n = x_\kappa \oplus n$ via voxel-wise addition.

\subsubsection{Exponentiation}

Gamma augmentation, or exponentiation, relates to gamma correction---a technique that bridges the nonlinear nature of brightness perception with the linear response of digital imaging systems. In our generative model, gamma augmentation randomizes image contrast by nonlinearly stretching or compressing the intensity range. First, we re-normalize the image to $[0, 1]$. Then, we sample a global exponent $\gamma \sim \U(a_\gamma, b_\gamma)$, $\gamma > 0$, and apply it voxel-wise:
\begin{equation}
    x_\gamma = \left[
    \frac{x_n - \operatorname{min}(x_n)}{\operatorname{max}(x_n) - \operatorname{min}(x_n)}
    \right] ^ \gamma.
\end{equation}

\subsubsection{Downsampling}

Together with the \nameref{sec:blur}, this step simulates a scan acquired at a lower resolution, upsampled to a higher resolution. First, we draw a downsampling factor $d_i \sim \U(1, b_d)$ for each axis $i \in \{1, 2, ..., N\}$ and downsample the image to reduce its size by these factors. Then, we upsample the result back to the original size to obtain a new image $x_d$. Both nearest-neighbor and linear interpolation are efficient choices. For the discrete-valued label map $s_x$, we use nearest-neighbor interpolation to avoid intermediate values when the downstream task requires downsampling of both $(x, s_x)$.

\subsubsection{Cropping}

We zero out a proportion of voxels at the edge of the FOV along randomized axes by multiplying the image voxel-wise with the binary mask $m$ from Section~\nameref{sec:mask}, yielding $x_m = x_d \odot m$. Such zero-intensity regions arise when preprocessing conforms images to a larger FOV. Depending on the size and location of the masked-out region, this step may partially crop the anatomy.

\subsubsection{Further corruptions}

In aggregate, the applied corruptions produce highly variable images for model training (Figure~\ref{fig:examples}). In order to enhance robustness to additional artifacts or preprocessing operations, incorporating further augmentations can be advantageous. For example, we may randomly clear image slices at random locations to simulate saturation effects. We might also set non-brain voxels to zero to simulate skull-stripping in preprocessing.
Additionally, it can be beneficial to apply more aggressive corruptions such as downsampling, smoothing, and cropping only some of the time.

With the generative model in place, we will now explore practical aspects of synthesis-driven DL.

\section{Practical discussion}
\label{sec:discussion}

The following sections outline key considerations for integrating the generative model into the DL stack and training networks with it, from a method selection framework to troubleshooting common issues.

\subsection{Method selection guidance}
\label{sec:selection}

Users seeking to choose between conventional and synthetic training need to consider their specific task, data---and to a lesser extent---computational resources. The main deciding factor is the task. Generally, domain randomization is appropriate for unsupervised problems that do not require ground-truth labels, such as registration with an overlap loss. Similarly compatible are supervised tasks whose supervision is invariant or equivariant with the transforms applied to label maps or images. For segmentation, for example, ground-truth label maps are invariant to additive image noise and equivariant with elastic augmentation. Synthesis-based training is also appropriate for tasks in which the quantity of interest can be derived from the image or label map. A counterexample is brain age prediction, as it is unclear how elastic augmentation may change the ground-truth age. Unless users only apply affine transforms or carefully explore \nameref{sec:ranges}, conventional training is more appropriate.
Domain randomization is an excellent choice for robust processing toolboxes. However, if the goal is to explore network architectures or solve a specific problem quickly, especially when generalization across a wide distribution of data is not required, conventional training is almost always preferable. Similarly, domain randomization is less suitable for data-driven analyses and exploratory DL aiming to characterize an empirical distribution~\cite{abukmeil2021survey,yang2024identifying}.

Another important factor is training data availability. A key advantage of synthesis-driven training is the ability to achieve competitive performance with few label maps. While reasonable performance is attainable with 5--10 subjects~\cite{billot2023synthseg}, some state-of-the-art methods use only about 100~\cite{hoopes2022synthstrip,hoffmann2024anatomy}.
Adding more label maps can increase accuracy, albeit with diminishing returns~\cite{billot2023synthseg}. In contrast, conventional DL often relies on datasets with several thousand subjects~\cite{balakrishnan2019voxelmorph}. Domain randomization is a suitable option when training images or labels are sparse. Manual labeling may be required to learn segmentation of new structures, but there are other ways to obtain labels of surrounding structures for synthesis (Section~\nameref{sec:labels})---labeling accuracy is largely irrelevant in synthesis-based training, since the generated images match the labels by construction.

Computational resources are typically not a limiting factor. As for any DL task, training a performant neural network requires a recent GPU. Running the generative model on the same device is efficient (Section~\nameref{sec:integration}); the added memory requirements are comparatively low as the model lacks trainable parameters. Some tasks benefit from increasing model capacity (Section~\nameref{sec:architecture}. The associated increase in training time and memory usage can often be accommodated by optimizing code, for example, by reducing the field of view, applying fewer convolutional filters at the highest resolution, operating at half resolution or training with mixed precision.

\subsection{Getting started}
\label{sec:start}

New adopters can quickly get started by building on publicly available implementations, which differ mainly in the order and details of the image corruption steps~\cite{billot2023synthseg,hoffmann2024anatomy}. Interactively experimenting with sampling ranges---such as the bias-field strength $\|B\| \sim \U(a_B, b_B)$---and generating examples in a Jupyter session will help users understand how the hyperparameters influence the image synthesis. 
In this context, setting both sampling bounds $(a_B, b_B)$ to the intended maximum value is a good habit allowing to visualize the strongest possible effect without generating many examples.
Table~\ref{tab:ranges} provides domain-randomization ranges to use as a starting point. Similar ranges have successfully been applied in multi- and cross-contrast MRI registration, segmentation, and skull-stripping of MRI, CT, and PET~\cite{hoopes2022synthstrip,billot2023synthseg,hoffmann2024anatomy}.

With the generative model set up, the next step is integrating it into the learning pipeline to train a task network. Unless tackling a completely new problem, begin by reproducing a known successful task and gradually expand from there.
Open-source demos are available at \url{https://w3id.org/synthmorph}, showcasing how to use publicly available code to synthesize images, train affine and deformable registration models, and use a domain-randomized model for CT-to-MRI registration. These demos run interactively in the browser and require no installation of additional software.

\subsection{Model inputs and outputs}

In conventional training, we sample an image $x$ and associated ground-truth quantity $y$ from the training set, pass $x$ through the task network, and compare the prediction $\hat{y}$ to $y$ in the loss function---before we compute its gradient with respect to the network weights to update these via backpropagation.
In the synthesis paradigm, we use a training set $\T$ of label maps. At each iteration, we sample a label map $s \in \T$ and use the generative model~$g$ to create a synthetic image-segmentation pair $(x, s_x)$ from $s$. The task network predicts $\hat{y}$ from $x$ as before---but what do we compare $\hat{y}$ to? In general, we derive the target quantity $y$ from $s_x$. For example, $y$ might be one-hot encoded brain structures for segmentation~\cite{billot2023synthseg} or an aggregate of brain structures for skull-stripping~\cite{hoopes2022synthstrip}. For unsupervised registration, we might generate two image-segmentation pairs, extract fixed brain labels $y$ and brain labels $\hat{y}$ moved by the estimated transform, and compare their one-hot overlap~\cite{hoffmann2021synthmorph}. Diagrams showing the information flow for these tasks are available in a recent review paper~\cite{gopinath2024synthetic}.

\begin{table}
    \centering
    \caption{Uniform domain-randomization starter ranges $[a, b]$, where SD abbreviates standard deviation. We list warp and bias field ranges assuming noise generation via \nameref{sec:gradient}.\label{tab:ranges}}
    \small
    \begin{tabular}{llrr}
    \toprule
    \textbf{Parameter} & \textbf{Unit} & $a$ & $b$ \\
    \midrule
    Translation $t$                       & mm       & $-30$ & 30 \\
    Rotation $r$                          & $^\circ$ & $-30$ & 30 \\
    Scaling $z$                           & \%       & 90    & 110 \\
    Shear $e$                             & \%       & 90    & 110 \\
    Warp strength $\|\mathbf{\phi}\|$     & mm       & 0     & 20 \\
    Warp control points $C$               &          & 2     & 16 \\
    Cropping proportion $p_m$             & \%       & 0     & 20 \\
    Label intensity mean $\mu$            & a.u.     & 0     & 1 \\
    Bias drop $\|B\|$                     & \%       & 0     & 50 \\
    Bias control points $C$               &          & 2     & 4 \\
    Image blurring SD $\sigma_\kappa$     & mm       & 0     & 2 \\
    Noise intensity SD $\sigma_n$         & \%       & 0     & 10 \\
    Gamma exponent $\gamma$               &          & 0.5   & 1.5 \\
    Downsampling factor $d$               &          & 1     & 4 \\
    \bottomrule
    \end{tabular}
\end{table}

\subsection{Efficient integration}
\label{sec:integration}

Novice users may be tempted to generate a static set of image-segmentation pairs to train a network as they would with real data~\cite{hendrickson2023bibsnet,zalevskyi2024improving}. We advise against this approach, as it undermines a key advantage of real-time synthesis: the ability to create a \textit{new} image with each invocation. Generating an effectively endless stream of training images maximizes the potential to learn generalizable features.

There are several ways to integrate real-time synthesis into the training pipeline. One option is to set up a combined model that takes label maps as input, synthesizes images from them, and processes these through the task model~\cite{billot2023synthseg,hoffmann2021synthmorph}. This setup requires extraction of the task model for validation and testing. A simpler method is calling separate models directly within the training loop.
Placing both synthesis and task models on the same GPU minimizes costly copy operations between host and device but requires sufficient GPU memory.
An alternative is to distribute synthesis and training over multiple processes or workers. For example, one implementation continuously generates training examples and writes them to disk using several CPU jobs, while a parallel GPU job reads the data for training, deleting samples after loading to minimize disk usage~\cite{hoopes2022synthstrip}. A recent project abstracts this approach into a Python package that orchestrates generation and training processes, efficiently streaming data through memory~\cite{doan2024scaling}.

\subsection{Randomization dimensions}

Randomizing every aspect of the training data is unnecessary, as allocating model capacity to handle heterogeneity that we can easily remove is counterproductive.
For example, simple min-max normalization standardizes the intensity range, while reorienting images based on header information avoids wasting capacity on learning all possible head orientations~\cite{hoffmann2024anatomy}.
In contrast, it is beneficial to identify characteristics of a new target distribution that the generative model may not synthesize yet. These characteristics should be randomized across a range that both covers \textit{and} exceeds the effects expected in real data. For instance, low-field MRI has a lower signal-to-noise ratio than standard MRI, and adjusting the randomization ranges accordingly can improve performance~\cite{laso2024quantifying}.
As discussed in Section~\nameref{sec:modeling}, successful implementations randomize not only the sampled values but also the sampling distributions: instead of adding noise from a normal distribution of fixed standard deviation, we also randomize the standard deviation. Similarly, we vary the spatial frequency and strength of nonlinear deformations.

\subsection{Randomization ranges}
\label{sec:ranges}

Randomizing image characteristics over wide ranges can improve model generalization by increasing the diversity of the training distribution, even though the synthesized images appear unrealistic. However, excessively wide ranges can make the task too difficult or force the network to allocate capacity to accommodate unnecessary variability. For example, random sampling of image intensities on a per-structure basis leads to unrealistic tissue contrasts but helps segmentation networks generalize beyond the intensity characteristics of specific modalities. Controlled deformation of input label maps improves registration performance, whereas excessive deformation can degrade it. This degradation usually occurs gradually: prior work showed validation accuracy on real data to vary smoothly within a relatively wide neighborhood of sampling-range optima~\cite{hoffmann2021synthmorph,fu2025learning}.

To ensure that the synthesis has a constructive impact on performance, we need to tailor the randomization ranges of Table~\ref{tab:ranges} to the specific task or target domain. Prior work on registration optimizes sampling ranges via grid search~\cite{hoffmann2021synthmorph}. This work fixes all but one synthesis hyperparameter and trains separate models for various values, initializing each with the same trained weights.
However, grid searches are computationally costly and challenging with tasks in which the applied corruptions may introduce concept shifts. For example, for brain age prediction from MRI, spatial-anatomical augmentation and applied corruptions may shift the brain age of the original anatomy in a way that is difficult to predict and control, potentially impinging on performance. A recent method explores the hyperparameter space in a more principled fashion: it learns optimal randomization ranges from a set of real, labeled images while simultaneously training a task network on data synthesized using these ranges~\cite{hu2024learn2synth}. Although the approach adds complexity, it reduces reliance on hand-crafted parameters.

\subsection{Label maps for synthesis}
\label{sec:labels}

Many publicly available neuroimaging datasets contain label maps that include brain and sometimes non-brain structures~\cite{hoopes2022synthstrip}. When label maps are unavailable, robust segmentation tools can derive them from the images~\cite{fischl2012freesurfer,billot2023synthseg}.
Although applications such as brain-specific registration only require brain labels to compute the loss, including non-brain labels enables synthesis of whole-head scans, improving network robustness across images with and without skull-stripping. Similarly, skull-stripping requires only two labels---brain and non-brain---but using only these labels for synthesis may limit generalization. We can create more complex image content by incorporating non-anatomical and artifactual structures segmented from images using simple methods such as k-means clustering or Gaussian mixture modeling~\cite{hoopes2022synthstrip,billot2023synthseg}.

Some works extend this approach by synthesizing the training label maps entirely, whether for segmentation~\cite{dey2024anystar,chollet2024neurovascular} or registration~\cite{hoffmann2021synthmorph}. This strategy eliminates the need for acquired training data and is well-suited for anatomy-agnostic tasks or situations where patch-wise processing allows the synthesis of simple anatomical features~\cite{chollet2024neurovascular}.

\subsection{Mixed-data training}

Combining real and synthetic training images is a powerful way of extending existing datasets, increasing network generality via domain randomization while including domain-specific knowledge. When label maps are available for the real data, alternating between real and synthetic enables use of the same loss function to avoid introducing a weighting problem~\cite{hendrickson2023bibsnet,zalevskyi2024improving}. When an additional loss term is required, such as an unsupervised image-similarity term for registration~\cite{balakrishnan2019voxelmorph}, we need to balance the loss function. Common approaches range from simple grid search and normalization by loss term magnitude to more principled methods like weighting by learned loss uncertainty or adaptive gradient norm balancing.

\subsection{Model architecture and capacity}
\label{sec:architecture}

Synthesis-driven training is compatible with networks of any architecture. In our experience, fully leveraging synthetic data can require larger networks, to capture the vast training distribution. For example, increasing network capacity leads to substantial gains in registration accuracy, constrained only by the available GPU memory~\cite{hoffmann2021synthmorph}. This work settles on a U-Net architecture with 10 convolutional layers of 256 filters each. In contrast, prior work on unsupervised registration, which the study builds upon, achieves state-of-the-art performance with a similar architecture using only 16 to 32 convolutional filters per layer~\cite{balakrishnan2019voxelmorph}.
However, the earlier work focuses on within-contrast registration of T1-weighted MRI, whereas networks trained with synthetic data tend to generalize well across MRI sequences and to cross-contrast registration.
A downside of larger networks is long training times---from a week for 3D segmentation to a month or longer for very large registration networks---making pilot experiments essential.

\subsection{Pilot experiments}
\label{sec:pilot}

Developing on a computationally reduced problem accelerates progress and helps eliminate bugs early, before switching to a full setup. Additionally, parameter sweeps become faster or more precise if hyperparameters transfer from the reduced to the full problem. We therefore recommend starting in 2D or at reduced resolution, where experiments often yield early results within an hour, and where models fully train overnight. For example, halving the resolution of 3D data cuts memory and processing demands by nearly 90\%. Synthesis hyperparameters generally transfer if adjusted for voxel size~\cite{hoffmann2024anatomy}---voxel-based ranges like blurring ($[a_\kappa, b_\kappa]$, Section~\nameref{sec:blur}) should double when moving from half to full resolution. While 2D experiments avoid this conversion, they require models with configurable dimensionality. Training on patches is also an efficient option if large-scale context is expendable. However, this strategy usually implies patch-based inference since performance drops when inputs deviate from the training distribution, and it requires fusing patch-wise predictions. Upon returning to full 3D runtimes, progressive validation during training becomes essential.

\subsection{Validating domain shift mitigation}
\label{sec:validation}

Continuously monitoring performance on real validation data ensures that training achieves its intended effect and helps determine when to stop. We recommend using multiple small but diverse validation sets that span various image types to reveal performance differences across domains. It is also helpful to train an identical network on real data in parallel, in order to quantify the reduction in domain shift induced by domain randomization.
For tasks like segmentation and registration, Dice scores on label maps offer a simple and interpretable metric.

\begin{figure}
    \centering
    \includegraphics[width=\columnwidth]{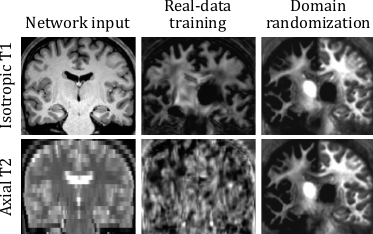}
    \caption{Features extracted from the last network layer for two scans of the same brain that differ in contrast and resolution. Domain randomization yields stable features, whereas training the same network on real T1-weighted data does not. Adapted from prior work~\cite{billot2023synthseg}.\label{fig:features}}
\end{figure}

A more direct way to assess domain-shift reduction is to measure the variability of network features in response to covariate shifts of interest, using image similarity metrics such as normalized cross-correlation or mean absolute difference. We expect features in the later layers of a deep network to become invariant to such shifts~\cite{billot2023synthseg,hoffmann2021synthmorph}. For example, a segmentation network should produce similar label maps for T1 and T2-weighted images of the same brain, and thus extract similar features in its final layers (Figure~\ref{fig:features}).

\subsection{Troubleshooting and failure modes}
\label{sec:troubleshooting}

Whenever adjusting a hyperparameter that influences the generative model, we advise generating 20--30 images and label maps to assess whether the change has the desired effect and does not introduce common failure modes: it is easy to unintentionally produce data that slow down convergence, cause frequent divergence, or lead to NaN loss values. Pitfalls include moving label maps out of the FOV, too much cropping when the anatomy is at the edge of the FOV, and applying image corruptions that set the entire image to zero. These failure modes can be readily identified by eye and corrected.

If the generation works as expected but the loss does not decrease, ensure that the model inputs and outputs are correct. The added complexity of the generative model can lead to incorrect tensor selections---for instance, passing the original label map $s$ to the loss when it should be $s_x$ after spatial augmentation. Visualizing inputs and outputs throughout the DL setup is often helpful. Another effective troubleshooting method is to write a single image-segmentation pair to disk and train the network on this one example, making feature extraction easy and convergence faster. If the loss does not decrease within this setup, there is likely a bug.

\subsection{Limitations of synthetic training}

Domain randomization can yield robust models that process a variety of neuroimaging data types without retraining. However, if the synthesis exclusively covers healthy anatomies or common pathologies, rarer patterns or atypical lesions may be handled suboptimally despite their clinical relevance. The limited size and diversity of many public datasets can introduce model bias when training on images synthesized from these data without sufficient spatial and anatomical augmentation. Bias can also occur when sampling rules within the generative model favor specific contrasts or shapes, leading to representation issues that impinge on balanced performance across populations.

Similarly, inadequate synthesis of complex real-world acquisition artifacts such as Gibbs ringing or radio-frequency inhomogeneities may cause networks to fail on scans that exhibit features outside the training distribution. Physiological noise from respiratory motion or pulsation is challenging to synthesize, yet it is often present in neuroimaging.
If the image generation fails to capture any one of these effects in training, networks risk overfitting to synthetic artifacts. This reality gap will limit the model's ability to generalize to real noise patterns and acquisition effects, thereby reducing performance.

Further limitations of the synthesis paradigm include a more abstract and complex setup, as well as longer training times. Users may need to adjust network capacity to handle the increased variability, select synthesis and augmentation ranges appropriate for the task, and efficiently allocate computational resources. Finally, domain randomization is not compatible with every DL task (Section~\nameref{sec:selection}).

\section{Conclusion and outlook}
\label{sec:conclusion}

Domain randomization is an emerging DL strategy in neuroimaging that trains generalizable networks using synthetic images. Despite impactful advances in core image-analysis tasks, many opportunities remain to explore and broaden the impact of domain-randomized training. Potential downstream applications include cortical surface reconstruction for morphometry, MRI bias-field correction, image quality estimation, and extensions to the frequency domain---k-space. To date, most works have focused on methodological development and image processing. Yet, synthesis-based strategies could prove transformative in earlier stages of the imaging pipeline---such as adaptive motion correction, autonomous patient-centric acquisition schemes, or undersampled reconstruction strategies.

Although domain randomization substantially enhances generalizability, it is unlikely to supplant real training data. On the contrary, these paradigms are fundamentally complementary. Learned corruption processes are a promising development that highlight their synergies: recent work has shown how parametric and residual degradations of synthetic images can be learned from real data to bridge reality gaps.
While this direction reduces the risk of wasting network capacity to learn unhelpful synthetic variability, understanding the trade-offs between model complexity, real data diversity, and the resulting generalizability remains an open research question.
Concurrently, cutting-edge diffusion and flow-based models have tremendous potential to enrich the image synthesis well beyond heuristic corruptions. These powerful generative learning approaches may condition the synthesis on more complex anatomical priors, spanning broad neurodiversity and rare pathological variants with greater fidelity. 

As neuroimaging and analysis continue to grow increasingly diverse and multimodal, scalable learning paradigms that complement real data to cover the high-dimensional space of anatomical, pathological, and technical heterogeneity will become ever more essential. This need is of particular importance for the development of large foundation models, as the availability and shareability of medical data remain comparatively restricted. The transformative impact that language models---specifically vision-language models---are having at large promises to reshape the neuroradiological landscape by enhancing diagnostic workflows through the synergistic integration of multimodal information, automated and adaptive reporting, as well as clinical decision support.

Looking ahead, domain randomization sits at an exciting intersection that combines domain-specific insight with principled generative modeling to achieve scalable learning. As generative technologies and foundation models evolve, we expect synthesis-based training to transition from handcrafted feature distributions towards data-driven, adaptive strategies---blurring the line between synthesis-based and real-world learning to pave a data-efficient pathway to general-purpose neuroimage analysis tools.

\section{Acknowledgment}

The work on this article was supported in part by the National Institute of Child Health and Human Development (R00 HD101553, R01 HD099846, R01 HD109436), the National Institute of Biomedical Imaging and Bioengineering (R01 EB033773), the National Institute on Aging (R01 AG064027), the National Institute of Neurological Disorders and Stroke (U24 NS135561), and the National Cancer Institute (R01 CA255479). The author maintains a consulting relationship with Neuro42, a company that did not have any involvement in the content of this work. He is grateful to Hanna Loetz for her unwavering support, without which this article would not have been possible.

\bibliographystyle{IEEEbib}
\bibliography{main}

\begin{thebibliography}{10}

\bibitem{fischl2012freesurfer}
Bruce Fischl,
\newblock ``{FreeSurfer},''
\newblock {\em NeuroImage}, vol. 62, no. 2, pp. 774--781, 2012.

\bibitem{malone2013miriad}
Ian~B Malone, David Cash, Gerard~R Ridgway, David~G MacManus, Sebastien Ourselin, Nick~C Fox, and Jonathan~M Schott,
\newblock ``{MIRIAD}—public release of a multiple time point {Alzheimer's MR} imaging dataset,''
\newblock {\em NeuroImage}, vol. 70, pp. 33--36, 2013.

\bibitem{althnian2021impact}
Alhanoof Althnian, Duaa AlSaeed, Heyam Al-Baity, Amani Samha, Alanoud~Bin Dris, Najla Alzakari, Afnan Abou~Elwafa, et~al.,
\newblock ``Impact of dataset size on classification performance: an empirical evaluation in the medical domain,''
\newblock {\em Applied Sciences}, vol. 11, no. 2, pp. 796, 2021.

\bibitem{zhou2022domain}
Kaiyang Zhou, Ziwei Liu, Yu~Qiao, Tao Xiang, et~al.,
\newblock ``Domain generalization: a survey,''
\newblock {\em IEEE Transactions on Pattern Analysis and Machine Intelligence}, vol. 45, no. 4, pp. 4396--4415, 2022.

\bibitem{gopinath2024synthetic}
Karthik Gopinath, Andrew Hoopes, Daniel~C Alexander, et~al.,
\newblock ``Synthetic data in generalizable, learning-based neuroimaging,''
\newblock {\em Imaging Neuroscience}, 2024.

\bibitem{wells1996adaptive}
William~M Wells, W~Eric~L Grimson, Ron Kikinis, and Ferenc~A Jolesz,
\newblock ``Adaptive segmentation of {MRI} data,''
\newblock {\em IEEE Transactions on Medical Imaging}, vol. 15, no. 4, pp. 429--442, 1996.

\bibitem{van2003unifying}
Koen Van~Leemput, Frederik Maes, Dirk Vandermeulen, et~al.,
\newblock ``A unifying framework for partial volume segmentation of brain {MR} images,''
\newblock {\em IEEE Transactions on Medical Imaging}, vol. 22, no. 1, pp. 105--119, 2003.

\bibitem{rueckert1999nonrigid}
Daniel R{\"u}ckert, Luke~I Sonoda, Carmel Hayes, Derek~LG Hill, Martin~O Leach, and David~J Hawkes,
\newblock ``Nonrigid registration using free-form deformations: application to breast {MR} images,''
\newblock {\em IEEE Transactions on Medical Imaging}, vol. 18, no. 8, pp. 712--721, 1999.

\bibitem{csurka2017domain}
Gabriela Csurka,
\newblock ``Domain adaptation for visual applications: a comprehensive survey,''
\newblock {\em arXiv preprint arXiv:1702.05374}, 2017.

\bibitem{wang2018deep}
Mei Wang and Weihong Deng,
\newblock ``Deep visual domain adaptation: A survey,''
\newblock {\em Neurocomputing}, vol. 312, pp. 135--153, 2018.

\bibitem{johnson2007adjusting}
W~Evan Johnson, Cheng Li, and Ariel Rabinovic,
\newblock ``Adjusting batch effects in microarray expression data using empirical bayes methods,''
\newblock {\em Biostatistics}, vol. 8, no. 1, pp. 118--127, 2007.

\bibitem{gretton2009covariate}
Arthur Gretton, Alex Smola, Jiayuan Huang, Marcel Schmittfull, et~al.,
\newblock ``Covariate shift by kernel mean matching,''
\newblock {\em Dataset shift in machine learning}, vol. 3, no. 4, pp. 5, 2009.

\bibitem{sun2017correlation}
Baochen Sun, Jiashi Feng, and Kate Saenko,
\newblock ``Correlation alignment for unsupervised domain adaptation,''
\newblock {\em Domain adaptation in computer vision applications}, pp. 153--171, 2017.

\bibitem{pan2019recent}
Zhaoqing Pan, Weijie Yu, Xiaokai Yi, Asifullah Khan, Feng Yuan, et~al.,
\newblock ``{Recent progress on generative adversarial networks (GANs): a survey},''
\newblock {\em IEEE Access}, vol. 7, pp. 36322--36333, 2019.

\bibitem{tzeng2017adversarial}
Eric Tzeng, Judy Hoffman, Kate Saenko, and Trevor Darrell,
\newblock ``Adversarial discriminative domain adaptation,''
\newblock in {\em IEEE Conference on Computer Vision and Pattern Recognition}, 2017, pp. 7167--7176.

\bibitem{shorten2019survey}
Connor Shorten and Taghi~M Khoshgoftaar,
\newblock ``A survey on image data augmentation for deep learning,''
\newblock {\em Journal of Big Data}, vol. 6, no. 1, pp. 1--48, 2019.

\bibitem{muller2023multimodal}
Gustav M{\"u}ller-Franzes, Jan~Moritz Niehues, Firas Khader, Soroosh~Tayebi Arasteh, Christoph Haarburger, Christiane Kuhl, Tianci Wang, Tianyu Han, Teresa Nolte, Sven Nebelung, et~al.,
\newblock ``A multimodal comparison of latent denoising diffusion probabilistic models and generative adversarial networks for medical image synthesis,''
\newblock {\em Scientific Reports}, vol. 13, no. 1, pp. 12098, 2023.

\bibitem{iglesias2023ready}
Juan~Eugenio Iglesias,
\newblock ``A ready-to-use machine learning tool for symmetric multi-modality registration of brain {MRI},''
\newblock {\em Scientific Reports}, vol. 13, no. 1, pp. 6657, 2023.

\bibitem{hoffmann2024anatomy}
Malte Hoffmann, Andrew Hoopes, Douglas~N Greve, Bruce Fischl, and Adrian~V Dalca,
\newblock ``Anatomy-aware and acquisition-agnostic joint registration with {SynthMorph},''
\newblock {\em Imaging Neuroscience}, 2024.

\bibitem{fu2025learning}
Jingru Fu, Adrian~V Dalca, Bruce Fischl, Rodrigo Moreno, and Malte Hoffmann,
\newblock ``{Learning accurate rigid registration for longitudinal brain MRI from synthetic data},''
\newblock in {\em IEEE International Symposium on Biomedical Imaging}. IEEE, 2025, pp. 1--5.

\bibitem{billot2023synthseg}
Benjamin Billot, Douglas~N Greve, Oula Puonti, Axel Thielscher, Koen Van~Leemput, Bruce Fischl, Adrian~V Dalca, et~al.,
\newblock ``{SynthSeg}: Segmentation of brain {MRI} scans of any contrast and resolution without retraining,''
\newblock {\em Medical Image Analysis}, vol. 86, pp. 102789, 2023.

\bibitem{hendrickson2023bibsnet}
Timothy~J Hendrickson, Paul Reiners, Lucille~A Moore, Jacob~T Lundquist, Begim Fayzullobekova, Anders~J Perrone, Erik~G Lee, Julia Moser, Trevor~KM Day, et~al.,
\newblock ``{BIBSNet}: A deep learning baby image brain segmentation network for {MRI} scans,''
\newblock {\em bioRxiv}, 2023.

\bibitem{zalevskyi2024improving}
Vladyslav Zalevskyi, Thomas Sanchez, Margaux Roulet, Jordina Aviles~Verdera, Jana Hutter, Hamza Kebiri, and Meritxell Bach~Cuadra,
\newblock ``Improving cross-domain brain tissue segmentation in fetal {MRI} with synthetic data,''
\newblock in {\em Medical Image Computing and Computer-Assisted Intervention}. Springer, 2024, pp. 437--447.

\bibitem{laso2024quantifying}
Pablo Laso, Stefano Cerri, Annabel Sorby-Adams, et~al.,
\newblock ``Quantifying white matter hyperintensity and brain volumes in heterogeneous clinical and low-field portable {MRI},''
\newblock in {\em IEEE International Symposium on Biomedical Imaging}. IEEE, 2024, pp. 1--5.

\bibitem{chollet2024neurovascular}
Etienne Chollet, Ya{\"e}l Balbastre, Chiara Mauri, et~al.,
\newblock ``{Neurovascular segmentation in sOCT with deep learning and synthetic training data},''
\newblock {\em arXiv preprint arXiv:2407.01419}, 2024.

\bibitem{hoopes2022synthstrip}
Andrew Hoopes, Jocelyn~S Mora, Adrian~V Dalca, et~al.,
\newblock ``{SynthStrip}: skull-stripping for any brain image,''
\newblock {\em NeuroImage}, vol. 260, pp. 119474, 2022.

\bibitem{kelley2024}
William Kelley, Nathan Ngo, Adrian~V. Dalca, Bruce Fischl, Lilla Z{\"o}llei, and Malte Hoffmann,
\newblock ``Boosting skull-stripping performance for pediatric brain images,''
\newblock in {\em IEEE International Symposium on Biomedical Imaging}. IEEE, 2024, pp. 1--5.

\bibitem{hoffmann2021synthmorph}
Malte Hoffmann, Benjamin Billot, Douglas~N Greve, Juan~Eugenio Iglesias, Bruce Fischl, and Adrian~V Dalca,
\newblock ``{SynthMorph}: learning contrast-invariant registration without acquired images,''
\newblock {\em IEEE Transactions on Medical Imaging}, vol. 41, no. 3, pp. 543--558, 2021.

\bibitem{dey2024anystar}
Neel Dey, Mazdak Abulnaga, Billot, et~al.,
\newblock ``{AnyStar}: domain randomized universal star-convex {3D} instance segmentation,''
\newblock in {\em IEEE/CVF Winter Conference on Applications of Computer Vision}, 2024, pp. 7593--7603.

\bibitem{hoffmann2023anatomy}
Malte Hoffmann, Andrew Hoopes, Bruce Fischl, and Adrian~V Dalca,
\newblock ``Anatomy-specific acquisition-agnostic affine registration learned from fictitious images,''
\newblock in {\em Medical Imaging 2023: Image Processing}. SPIE, 2023, vol. 12464, p. 1246402.

\bibitem{dey2024learning}
Neel Dey, Benjamin Billot, Hallee~E. Wong, Clinton Wang, Mengwei Ren, Ellen Grant, et~al.,
\newblock ``Learning general-purpose biomedical volume representations using randomized synthesis,''
\newblock in {\em International Conference on Learning Representations}, 2025, pp. 1--32.

\bibitem{iglesias2021joint}
Juan~Eugenio Iglesias, Benjamin Billot, Ya{\"e}l Balbastre, Azadeh Tabari, John Conklin, R~Gilberto Gonz{\'a}lez, Daniel~C Alexander, Polina Golland, Brian~L Edlow, Bruce Fischl, et~al.,
\newblock ``Joint super-resolution and synthesis of 1 mm isotropic {MP-RAGE} volumes from clinical {MRI} exams with scans of different orientation, resolution and contrast,''
\newblock {\em NeuroImage}, vol. 237, pp. 118206, 2021.

\bibitem{perlin1985image}
Ken Perlin,
\newblock ``An image synthesizer,''
\newblock {\em ACM Siggraph Computer Graphics}, vol. 19, no. 3, pp. 287--296, 1985.

\bibitem{arsigny2006log}
Vincent Arsigny, Olivier Commowick, Xavier Pennec, and Nicholas Ayache,
\newblock ``A log-euclidean framework for statistics on diffeomorphisms,''
\newblock in {\em Medical Image Computing and Computer-Assisted Intervention}. Springer, 2006, pp. 924--931.

\bibitem{sled1998nonparametric}
John~G Sled, Alex~P Zijdenbos, and Alan~C Evans,
\newblock ``A nonparametric method for automatic correction of intensity nonuniformity in {MRI} data,''
\newblock {\em IEEE Transactions on Medical Imaging}, vol. 17, no. 1, pp. 87--97, 1998.

\bibitem{abukmeil2021survey}
Mohanad Abukmeil, Stefano Ferrari, Angelo Genovese, Vincenzo Piuri, and Fabio Scotti,
\newblock ``A survey of unsupervised generative models for exploratory data analysis and representation learning,''
\newblock {\em Acm Computing Surveys}, vol. 54, no. 5, pp. 1--40, 2021.

\bibitem{yang2024identifying}
Zhijian Yang, Junhao Wen, Guray Erus, et~al.,
\newblock ``Identifying five dominant dimensions of neurodegeneration in brain aging through deep learning: correlations with clinical and genetic measures,''
\newblock in {\em Alzheimer's Association International Conference}. ALZ, 2024.

\bibitem{balakrishnan2019voxelmorph}
Guha Balakrishnan, Amy Zhao, Mert~R Sabuncu, John Guttag, and Adrian~V Dalca,
\newblock ``{Voxelmorph: a learning framework for deformable medical image registration},''
\newblock {\em IEEE Transactions on Medical Imaging}, vol. 38, no. 8, pp. 1788--1800, 2019.

\bibitem{doan2024scaling}
Mike Doan and Sergey Plis,
\newblock ``Scaling synthetic brain data generation,''
\newblock {\em IEEE Journal of Biomedical and Health Informatics}, 2024.

\bibitem{hu2024learn2synth}
Xiaoling Hu, Oula Puonti, Juan~Eugenio Iglesias, Bruce Fischl, and Yael Balbastre,
\newblock ``{Learn2Synth}: learning optimal data synthesis using hypergradients,''
\newblock {\em arXiv preprint arXiv:2411.16719}, 2024.

\end{thebibliography}

\end{document}